# Non-uniform mesh based FDM simulation of lid-driven cavity problem governed by N-S equations in stream function-vorticity formulation

Zirui Mao[*]

Department of Aerospace Engineering and Engineering Mechanics,
University of Cincinnati, Cincinnati, OH, 45219, USA

**Abstract:** In this paper, the driven cavity problem was solved using finite difference scheme in stream function-vorticity formulation. A variable grid is adopted to capture more details and information in the area nearby the wall. The Navier-Stokes equation is rewritten as a particular form suitable to the variable grids. In simulation, Reynolds number is set 100 as an example. The velocity, vorticity and streamline contour are produced by the CFD scheme developed in this paper and then are compared with those by Ghia *et. al.* (1982) to validate this numerical scheme. It shows that the numerical CFD scheme developed in this paper works very well for both uniform grids and variable grids. The numerical tests with different grids setting show that a) the variable grids have advantages in capturing the reversed flow and separation bubbles produced in the corners associated with a good efficiency, b) the numerical schemes with symmetric and dense grids gives a more accurate solution than those with non-symmetric and sparse grids, and c) both the vorticity and stream function have a better accuracy than velocity.

## 1. Introduction

The N-S equations in stream function-vorticity formulation is simple and straightforward to implement and advantageous in handling the flows with the existence of vorticity. Typically, the variables gradients in the N-S equations are approximated by the finite difference schemes with a uniform grid. Compare to the regular grid, the variable nodes separations can always produce a better computation efficiency through coarsening the mesh in the less interested region. Till now, there is no related literature is found to investigate the effect of variable grid on the numerical results yet. Therefore, this work studies this topic.

---

[*] Corresponding author: maozu@mail.uc.edu.

The lid-driven cavity flow is a classic validation example in computational fluid dynamics to validate one numerical method. In this study, it is used to verify the proposed FDM with a N-S equation in stream function-vorticity formulation with a variable spacing ratio.

The cavity flow problem is described in the following figure 1. Basically, there is a constant velocity across the top of the cavity which creates a circulating flow inside. To simulate this flow a constant velocity boundary condition is applied to the lid, while the other three walls obey the no slip condition. Different Reynolds numbers give different results, so in this study Re=100 is adopted as an example. At high Reynolds numbers we expect to see a more interesting result with secondary circulation zones forming in the corners of the cavity, but in this paper this will not show up.

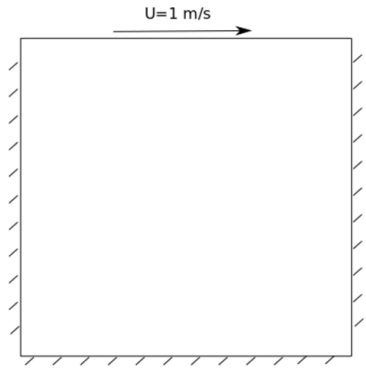

Fig. 1 Cavity flow problem

*Governing equation*

In 2D incompressible flow, the Navier-Stokes equations in stream function-vorticity formulation can be rewritten to be the following two equations:

$$\omega_t + u\omega_x + v\omega_y = \frac{1}{Re}\left(\omega_{xx} + \omega_{yy}\right) \quad (1)$$

$$\psi_{xx} + \psi_{yy} = -\omega \quad (2)$$

Where,

$$\omega = v_x - u_y \quad (3)$$

$$u = \psi_y; \quad v = -\psi_x \quad (4)$$

## 2. Finite difference scheme and solver algorithm

### 2.1 Discretization of the governing equations

In this paper, finite difference method is applied to discretize the governing equations above and create a system of equations that can be solved numerically associated with variable grids. Finite differences scheme is used to approximate the derivatives at each point using its surrounding points. This is shown in figure 2. By doing the discreated equation can be obtained for each node inside the cavity.

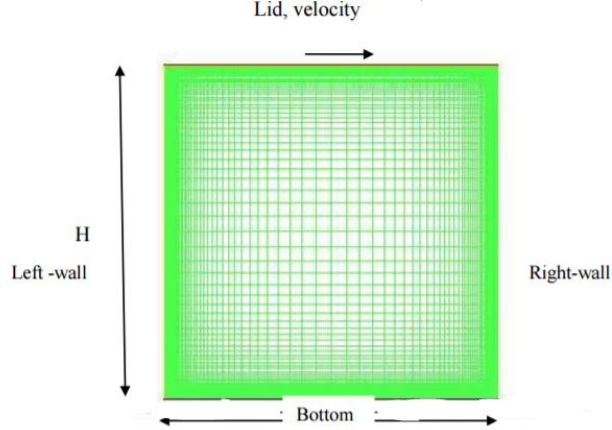

Fig. 2 Example cavity finite difference grid

The N-S equations in stream function-vorticity formulation represented by equations (1)-(2) contain first and second derivatives in spatial dimension, and a first derivative in time. Since the variable grids are adopted in this paper, the derivatives approximations have the following expression and accuracy:

$$\omega_t = \frac{\omega_{i,j}^{k+1} - \omega_{i,j}^{k}}{\Delta t} + O(\Delta t) \tag{5}$$

$$\omega_x = \frac{\omega_{i+1,j}^{k} - \omega_{i-1,j}^{k}}{\Delta x_i + \Delta x_{i-1}} + O(\Delta x) \tag{6}$$

$$\omega_y = \frac{\omega_{i,j+1}^{k} - \omega_{i,j-1}^{k}}{\Delta y_j + \Delta y_{j-1}} + O(\Delta y) \tag{7}$$

$$\omega_{xx} = \frac{2\left(\omega_{i+1,j}^{k}\Delta x_{i-1} - \omega_{i,j}^{k}\left(\Delta x_i + \Delta x_{i-1}\right) + \omega_{i-1,j}^{k}\Delta x_i\right)}{\Delta x_i \Delta x_{i-1}\left(\Delta x_i + \Delta x_{i-1}\right)} + O(\Delta x) \tag{8}$$

$$\omega_{yy} = \frac{2\left(\omega_{i,j+1}^{k}\Delta y_{j-1} - \omega_{i,j}^{k}\left(\Delta y_j + \Delta y_{j-1}\right) + \omega_{i,j-1}^{k}\Delta y_j\right)}{\Delta y_j \Delta y_{j-1}\left(\Delta y_j + \Delta y_{j-1}\right)} + O(\Delta y) \tag{9}$$

$$\psi_{xx} = \frac{2\left(\psi_{i+1,j}^{k}\Delta x_{i-1} - \psi_{i,j}^{k}\left(\Delta x_{i} + \Delta x_{i-1}\right) + \psi_{i-1,j}^{k}\Delta x_{i}\right)}{\Delta x_{i}\Delta x_{i-1}\left(\Delta x_{i} + \Delta x_{i-1}\right)} + O(\Delta x) \tag{10}$$

$$\psi_{yy} = \frac{2\left(\psi_{i,j+1}^{k}\Delta y_{j-1} - \psi_{i,j}^{k}\left(\Delta y_{j} + \Delta y_{j-1}\right) + \psi_{i,j-1}^{k}\Delta y_{j}\right)}{\Delta y_{j}\Delta y_{j-1}\left(\Delta y_{j} + \Delta y_{j-1}\right)} + O(\Delta y) \tag{11}$$

Where, $k$ represents the time index, and

$$\Delta x_{i} = x_{i+1} - x_{i}; \qquad \Delta y_{j} = y_{j+1} - y_{j} \tag{12}$$

The equations (5)-(12) are inserted into equation (1) and (2) to get the numerical scheme for this cavity flow problem. In specific, equation (1) is solved firstly to calculate the vorticity in next step. Then the new vorticity is used to calculate the stream function by solving equation (2).

**2.2 Solving the vorticity equation in time**

In this paper, the explicit Euler scheme is applied in time discretization as shown in equation (5). After substituting equation (5) into equation (2), we can get the following formulation for the calculation of vorticity in next step based on the current one:

$$\omega_{i,j}^{k+1} = \omega_{i,j}^{k} + \Delta t\left(\frac{1}{\text{Re}}\left(\omega_{xx} + \omega_{yy}\right) - \psi_{y}\omega_{x} + \psi_{x}\omega_{y}\right) \tag{13}$$

Equations (5)-(12) are used to approximate the derivatives about x and y in equation (13) above.

Although the stream function-vorticity formulation for N_S equation have many advantages in solving 2D incompressible problems, one of obvious difficulties is the more complicated boundary condition setting. As what we learnt in the class, the stream function boundary condition is very simple and straightforward, but the boundary condition for vorticity is much more complicated. Here, Taylor series is adopted to derive suitable boundary conditions for vorticity.

Using the top wall as an example, from equation (1) we know that:

$$\left(\frac{\partial^{2}\psi}{\partial y^{2}}\right)_{i,N} = -\omega_{i,N} \tag{14}$$

The second derivative in x direction vanishes here because the stream function is constant along the top wall. Considering Taylor series,

$$\psi_{i,N-1} = \psi_{i,N} - \left(\frac{\partial \psi}{\partial y}\right)_{i,N}\Delta y_{N-1} + \left(\frac{\partial^{2}\psi}{\partial y^{2}}\right)_{i,N}\frac{\Delta y_{N-1}^{2}}{2} + \ldots \tag{15}$$

By rearranging equation (14) and (15), the boundary condition for vorticity in the top can be obtained:

$$\omega_{i,N} = -2\frac{\psi_{i,N-1}}{\Delta y_{N-1}^2} - \frac{2}{\Delta y_{N-1}} \tag{16}$$

Similarly, the boundary condition for vorticity in the bottom:

$$\omega_{i,1} = -2\frac{\psi_{i,2}}{\Delta y_1^2} \tag{17}$$

Boundary condition for vorticity in the left:

$$\omega_{1,j} = -2\frac{\psi_{2,j}}{\Delta x_1^2} \tag{18}$$

Boundary condition for vorticity in the right:

$$\omega_{M,j} = -2\frac{\psi_{M-1,j}}{\Delta x_{M-1}^2} \tag{19}$$

### 2.3 Solving the stream function

After predicting the vorticity at the next time step, the new vorticity is used to solve equation (2) in order for calculating the stream function at the next time step.

By substituting equation (5)-(12) into equation (2), the stream function can be obtained using the following form:

$$\psi_{i,j} = \frac{C_x \psi_{i+1,j} \Delta x_{i-1} + C_x \psi_{i-1,j} \Delta x_i + C_y \psi_{i,j+1} \Delta y_{j-1} + C_y \psi_{i,j-1} \Delta y_j + \omega_{i,j}}{C_x (\Delta x_i + \Delta x_{i-1}) + C_y (\Delta y_j + \Delta y_{j-1})} \tag{20}$$

Where

$$C_x = \frac{2}{\Delta x_i \Delta x_{i-1} (\Delta x_i + \Delta x_{i-1})}; \quad C_y = \frac{2}{\Delta y_j \Delta y_{j-1} (\Delta y_j + \Delta y_{j-1})} \tag{21}$$

The derivation process is not shown here, because it is very simple and direct by inserting equation (5)-(12) into equation (2) and rearranging it.

For the boundary condition of stream function, considering the no flux property in all the boundaries and relationship between the stream function and velocity represented by equation (4), the stream function in all boundaries is constant and consistent. In this paper, it is set as zero for simplicity.

### 2.4 Convergence criterion

The residual criterions for streamfunction and vorticity are adopted to evaluate the convergence condition of calculation process. They are respectively given as

$$RES_\psi = \max\left(\left|\psi_{i,j}^{k+1} - \psi_{i,j}^k\right|\right) \qquad (22)$$

$$RES_\omega = \max\left(\left|\omega_{i,j}^{k+1} - \omega_{i,j}^k\right|\right) \qquad (23)$$

In our calculation, for all Reynolds numbers, we considered that convergence was achieved when both $RES_\psi < 10^{-17}$ and $RES_\omega < 10^{-15}$ was satisfied, which means the stream function and vorticity variables are accurate to $16^{th}$ and $14^{th}$ digit accuracy respectively at a grid point and even more accurate at the rest of the grids. Such a low value was chosen to ensure the accuracy of the solution and to capture the separation bubbles in the corners as best as possible.

### 2.5 Solver algorithm

The calculation process can be listed as following:
1. Specify the initial stream function and vorticity at all points;
2. Loop through all interior nodes and calculate the time derivative and use explicit Euler scheme to predict the vorticity at each point at the next time step (k+1); in this study, time step dt is set as 0.001 sec; At boundary nodes, calculate the vorticity using Taylor series approximation;
3. Loop through all the interior nodes and solve stream function equation to predict the stream function; Keep the stream function at all boundaries the same ($\psi$=0);
4. Check for convergence (based on the maximum relative vorticity, streamfunction and pressure criterions in sequent two steps);
5. If converged the simulation is done, plot the interested variables, such as velocity, pressure, vorticity and stream function and so on. If not converged, go back to step 2 and continue forward solving in time until converging.

## 3. Validation and discussions of numerical results

### 3.1 Validation of the numerical scheme

The current CFD scheme is firstly validated by comparing its results with experimental result by Ghia et. al (1982) with Re=100 and 129*129 uniform grids.

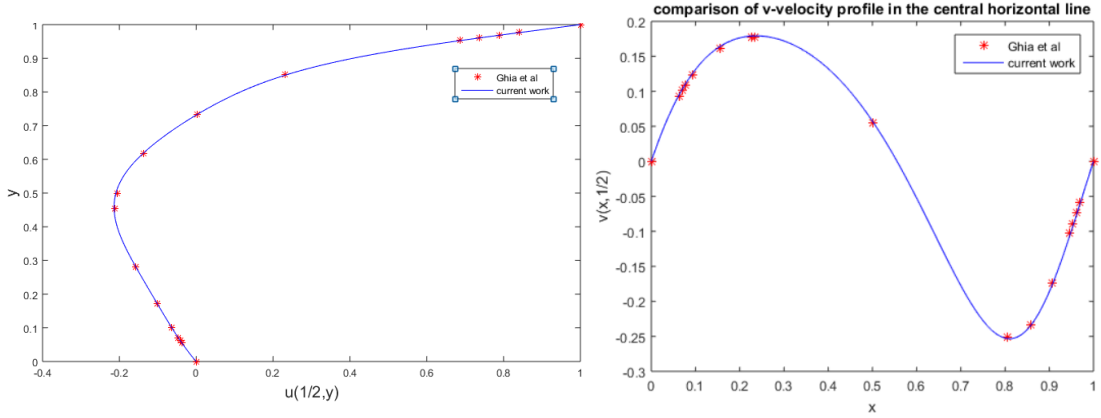

Fig. 3 Comparison of u-velocity profile along a vertical line passing though the geometry center (left) and v-velocity in the horizontal line passing through the geometry center with results from Ghia et al.

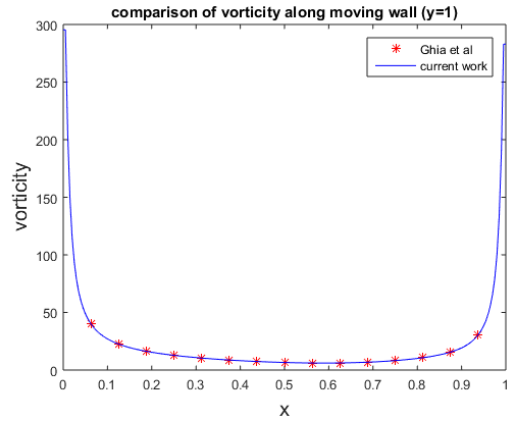

Fig. 4 Comparison of vorticity along moving wall (y=1) with Ghia et al. results at Re=100

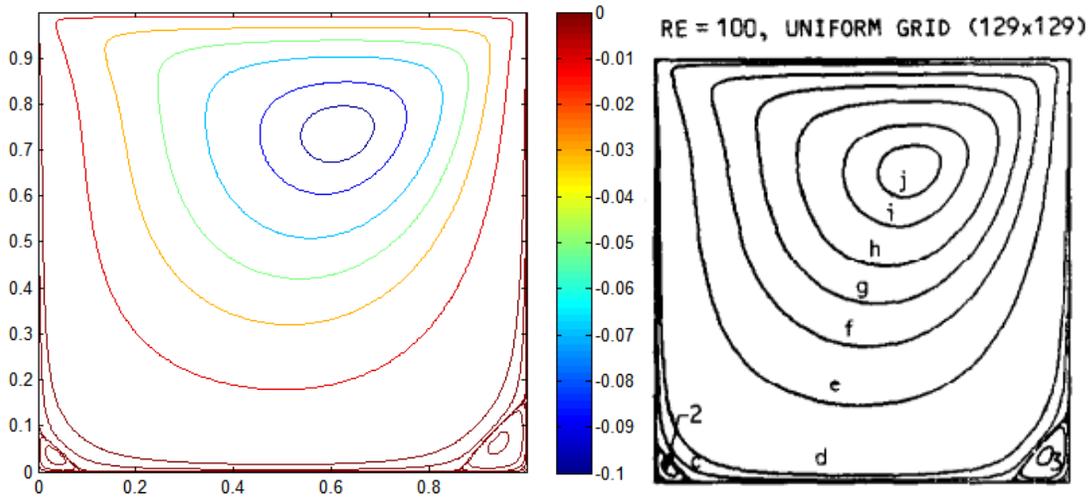

Fig 5. Comparison of stream function contour with Ghia et al results with Re=100 (the stream line value is set as what Ghia et al did shown in Fig. 7)

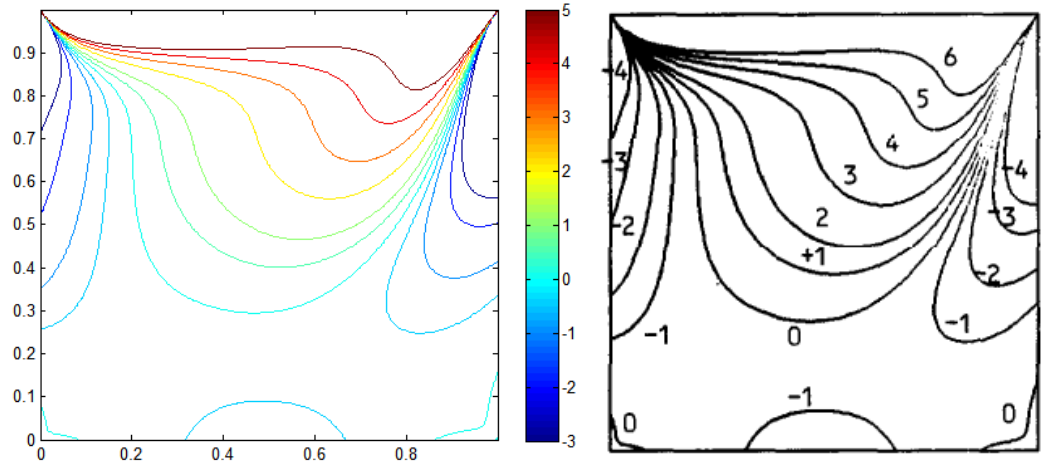

Fig. 6 Comparison of vorticity contour with Ghia et al results with Re=100 (the vorticity values are set as figure 7 showing)

|  | Stream function |  |  | Vorticity |  |
|---|---|---|---|---|---|
| Contour letter | Value of $\psi$ | Contour number | Value of $\psi$ | Contour number | Value of $\omega$ |
| a | $-1.0 \times 10^{-10}$ | 0 | $1.0 \times 10^{-8}$ | 0 | 0.0 |
| b | $-1.0 \times 10^{-7}$ | 1 | $1.0 \times 10^{-7}$ | ±1 | ±0.5 |
| c | $-1.0 \times 10^{-5}$ | 2 | $1.0 \times 10^{-6}$ | ±2 | ±1.0 |
| d | $-1.0 \times 10^{-4}$ | 3 | $1.0 \times 10^{-5}$ | ±3 | ±2.0 |
| e | $-0.0100$ | 4 | $5.0 \times 10^{-5}$ | ±4 | ±3.0 |
| f | $-0.0300$ | 5 | $1.0 \times 10^{-4}$ | 5 | 4.0 |
| g | $-0.0500$ | 6 | $2.5 \times 10^{-4}$ | 6 | 5.0 |
| h | $-0.0700$ | 7 | $5.0 \times 10^{-4}$ |  |  |
| i | $-0.0900$ | 8 | $1.0 \times 10^{-3}$ |  |  |
| j | $-0.1000$ | 9 | $1.5 \times 10^{-3}$ |  |  |
| k | $-0.1100$ | 10 | $3.0 \times 10^{-3}$ |  |  |
| l | $-0.1150$ |  |  |  |  |
| m | $-0.1175$ |  |  |  |  |

Fig. 7 Values of streamline contours in Figure 5 and 6

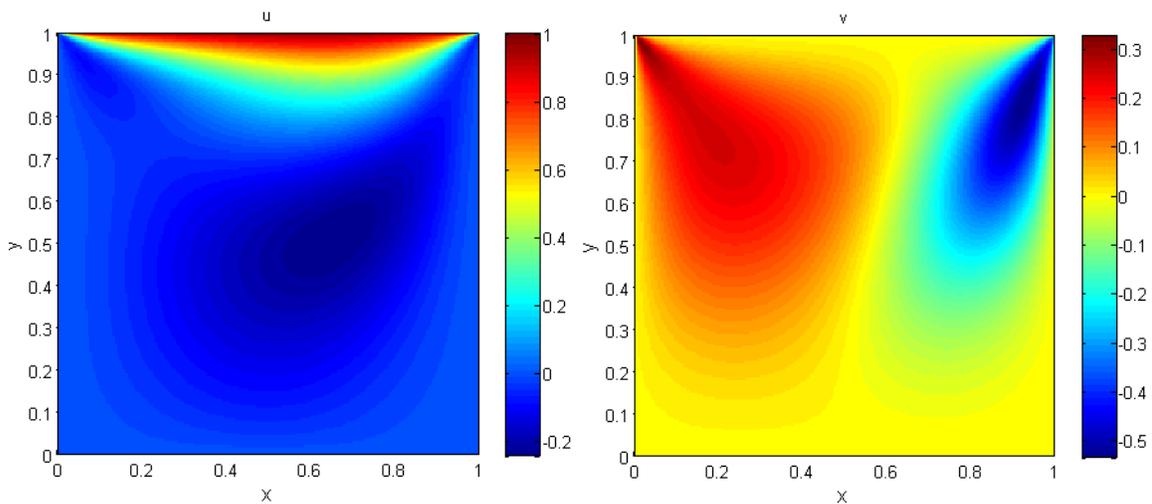

Fig. 8 Velocity field in x (left) and y (right) direction with 129*129 nodes and Re=100

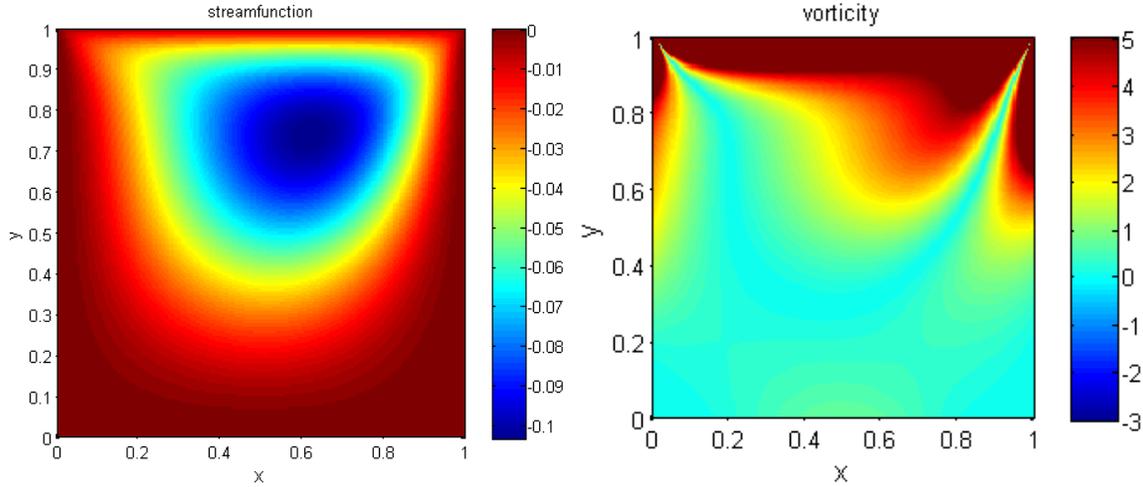

Fig. 9 Stream function (Left) and vorticity (Right) fields from the current scheme

From Figure (3-7), we can easily observe that the current CFD scheme matches very well with the results from Ghia et al (1982), which means that this numerical scheme and computer codes is reliable and accurate in simulating the driven cavity flow.

Figure 8 and 9 gives the velocity, stream function and vorticity field obtained from this current FDM developed in this paper.

### 3.2 Effects of variable grids on the numerical results

In this part, the effect of variable grids on the numerical solutions is investigated. In specific, the symmetric and non-symmetric grids are adopted respectively as shown in table 1. Figure 10 shows how the grids look like.

In simulations, the maximum ratio of length to width of rectangles are controlled not larger than 10 for a better numerical result. Hence, the ratio of $x_+/x_-$ and $y_+/y_-$ is set as 1.1 in scheme II and as 1.05 in scheme III, IV and V when 51*51 grid points are adopted in the following simulations.

Figure 11-15 show the comparison of velocity and vorticity with Ghia's results. Figure 11 shows the u-velocity in middle vertical line, and figure 12 shows the v-velocity in middle horizontal line, both of which are compared with Ghia's result. Figure 13 and 14 provide the stream line and vorticity contour. In figure 15, the vorticity in the top wall is compared with that from Ghia's work.

Table 1. 5 schemes with different settings and symmetry of grids using 51*51 nodes

|  | X symmetry | Y symmetry | Ratio $x_+/x_-$ | Ratio $y_+/y_-$ |
|---|---|---|---|---|
| Scheme I | Y | Y | 1.0 | 1.0 |
| Scheme II | Y | Y | 1.1　　left half<br>1/1.1 right half | 1.1　　top half<br>1/1.1 below half |
| Scheme III | Y | N | 1.0 | 1/1.05 |
| Scheme IV | N | Y | 1/1.05 | 1.0 |
| Scheme V | N | N | 1/1.05 | 1/1.05 |

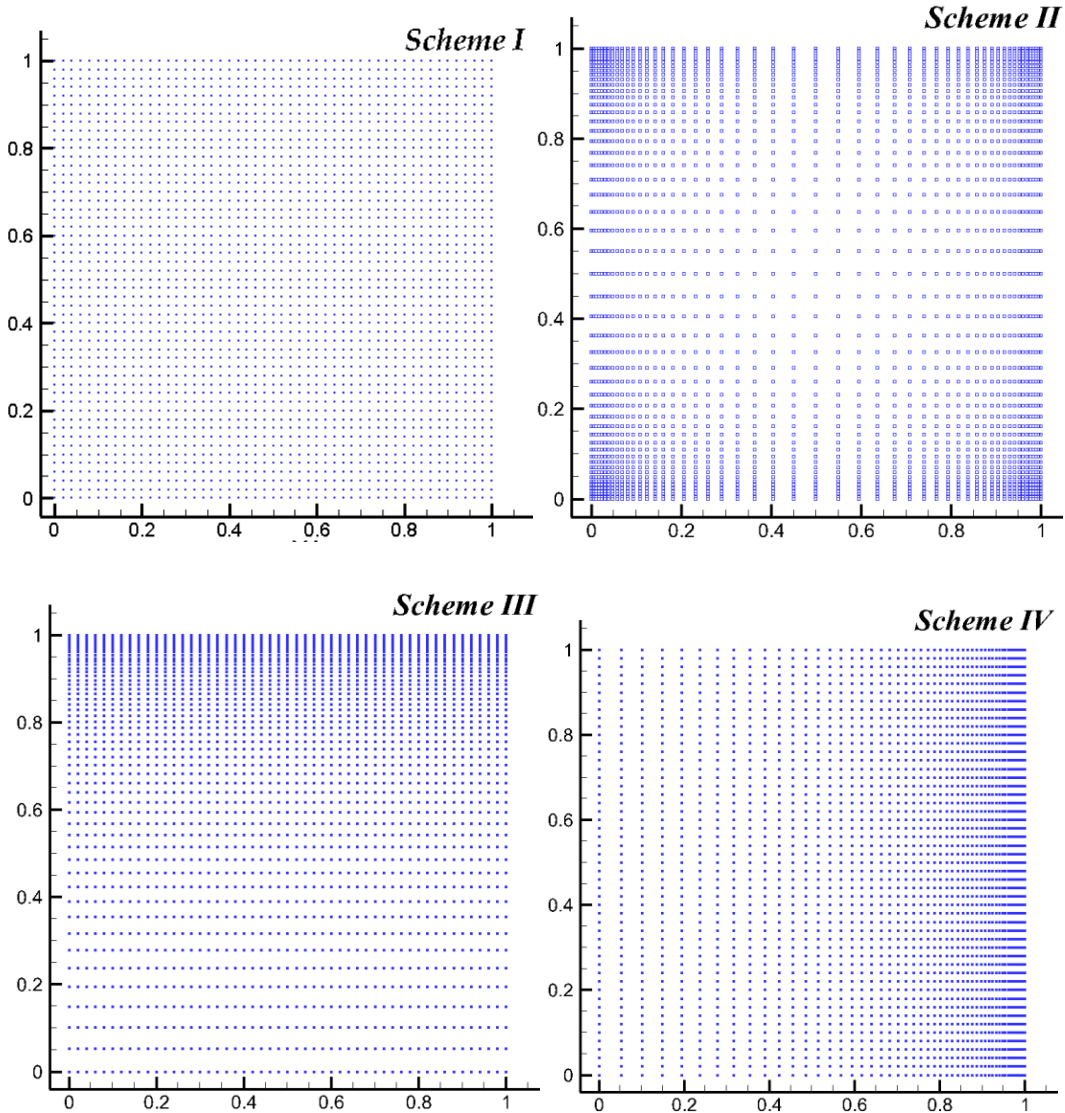

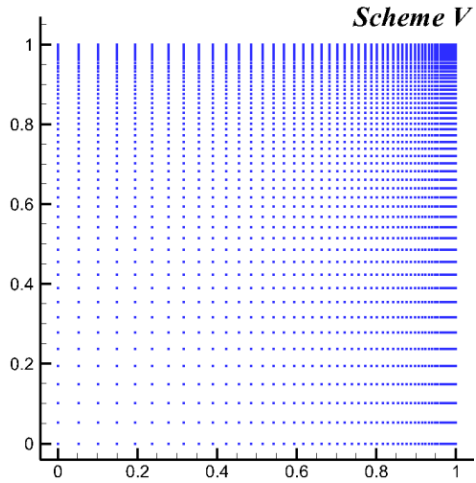

Fig. 10 Computational grids of 5 different schemes as shown in table 1

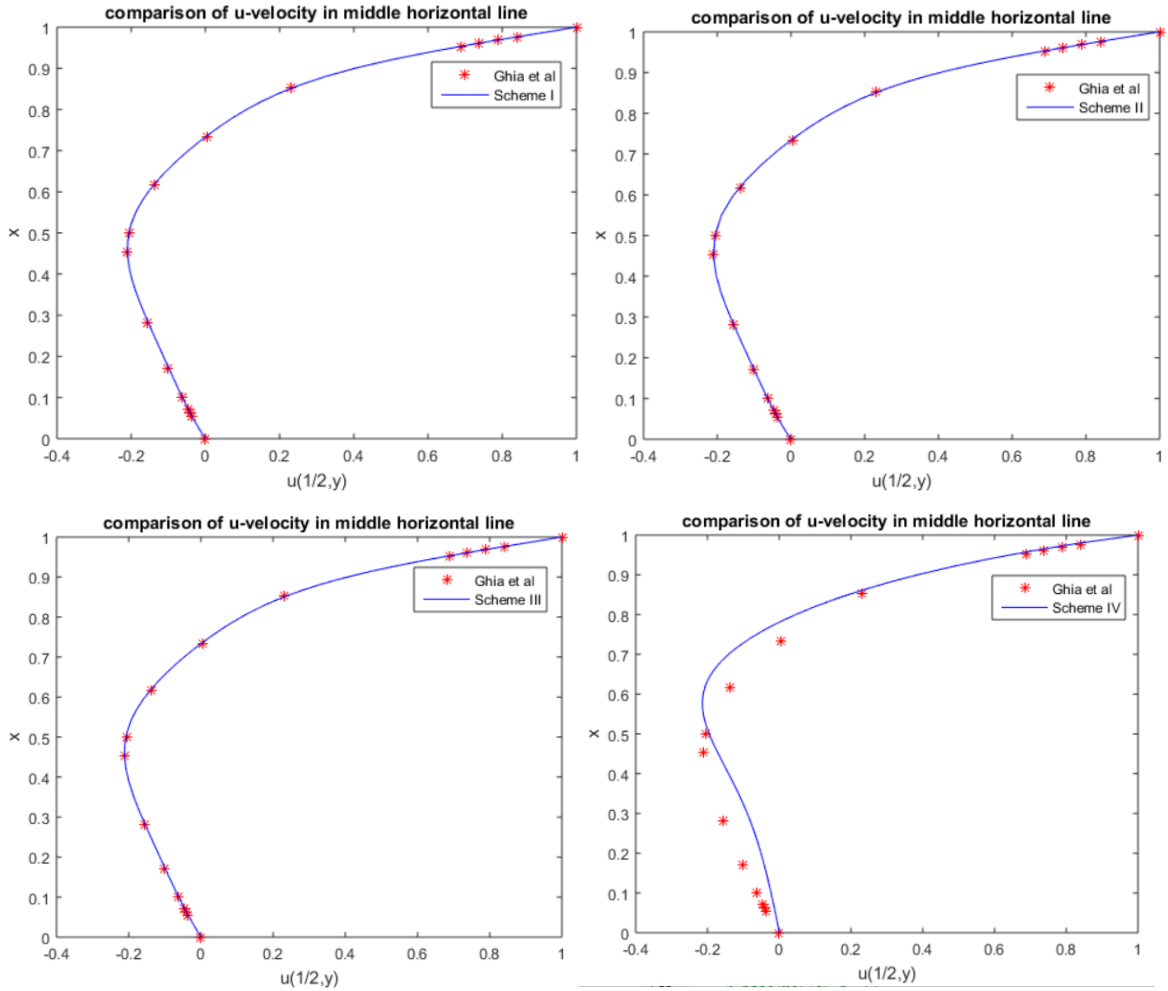

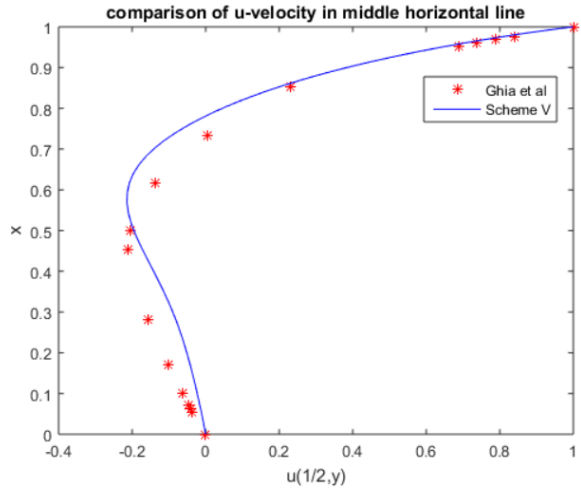

Fig. 11 Comparison of u-velocity in middle vertical line with Ghia's result

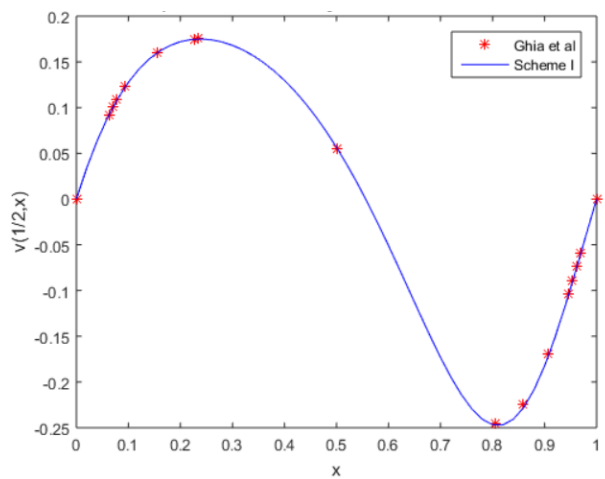
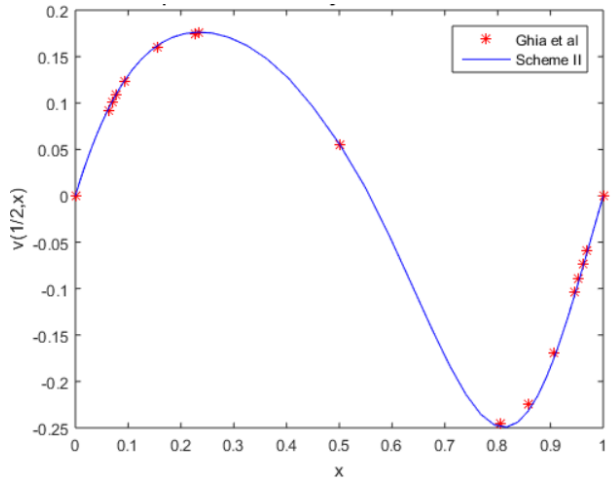
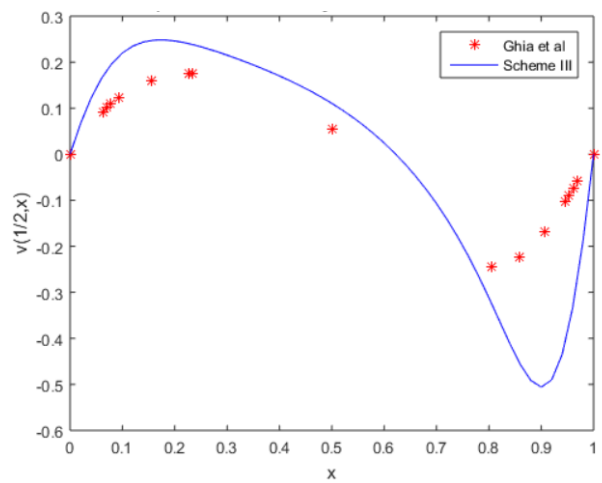
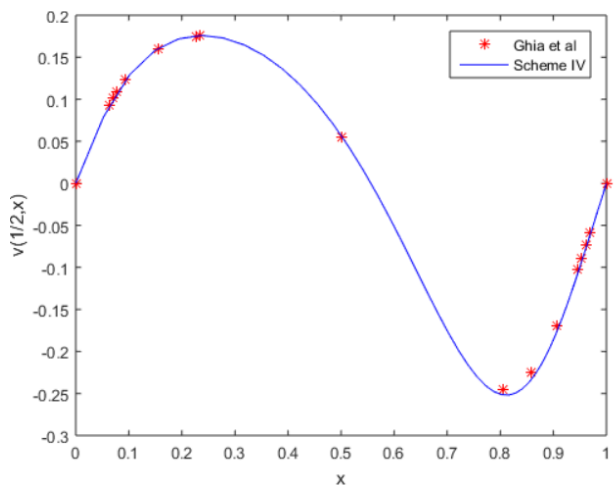

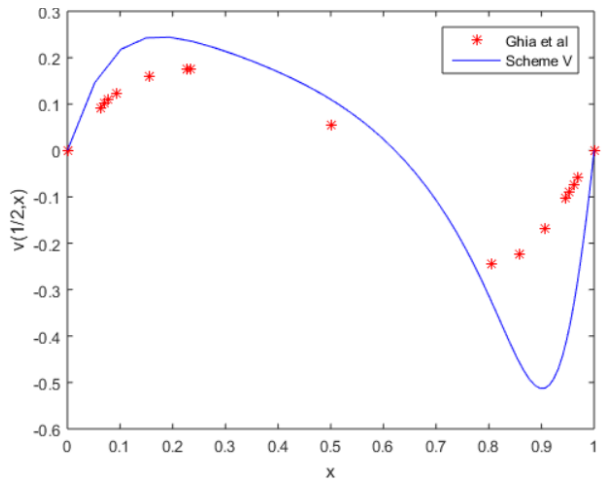

Fig. 12 comparison of v-velocity in middle horizontal line with Ghia's results

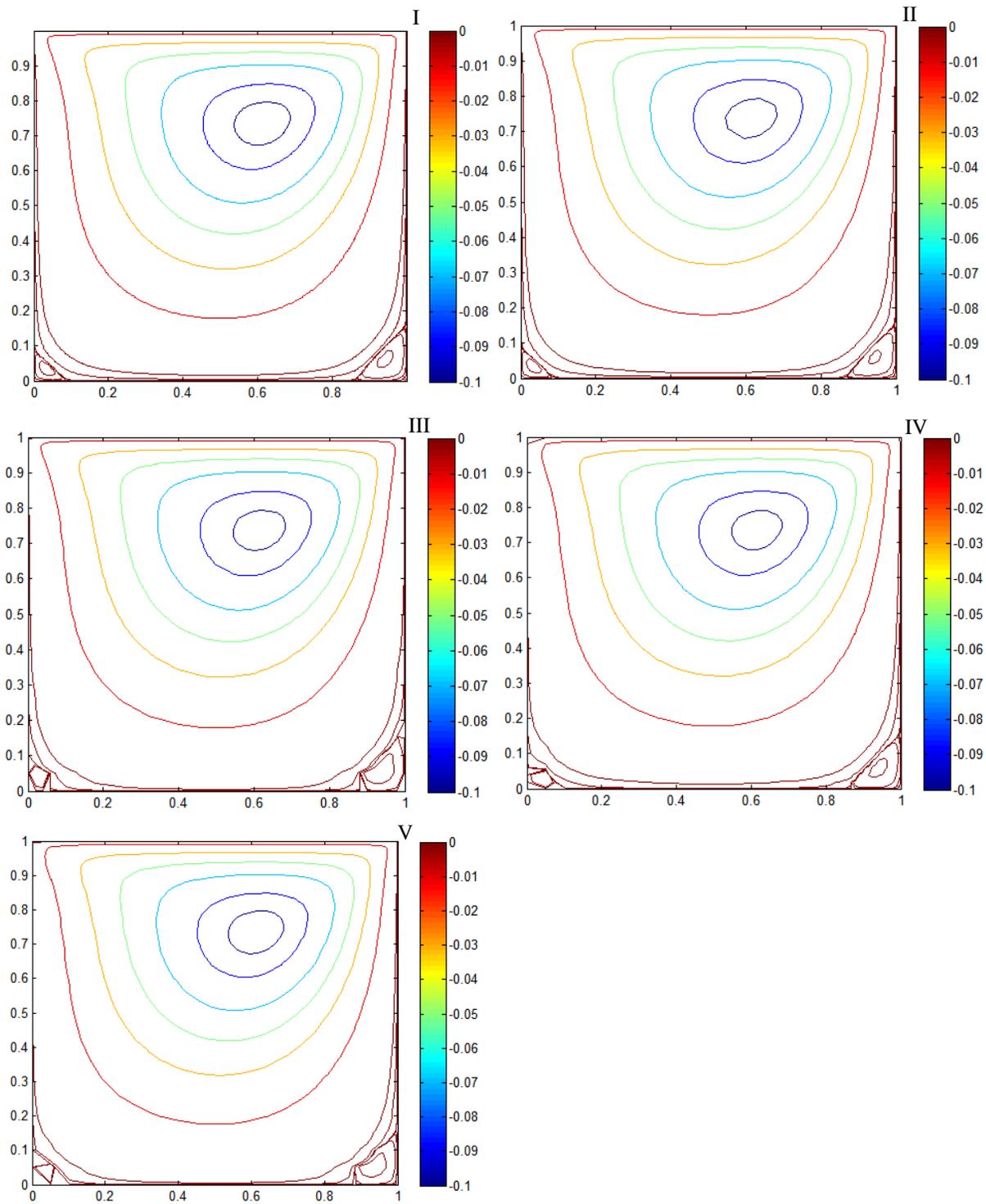

Fig. 13 Stream line contour using different schemes as shown in table 1

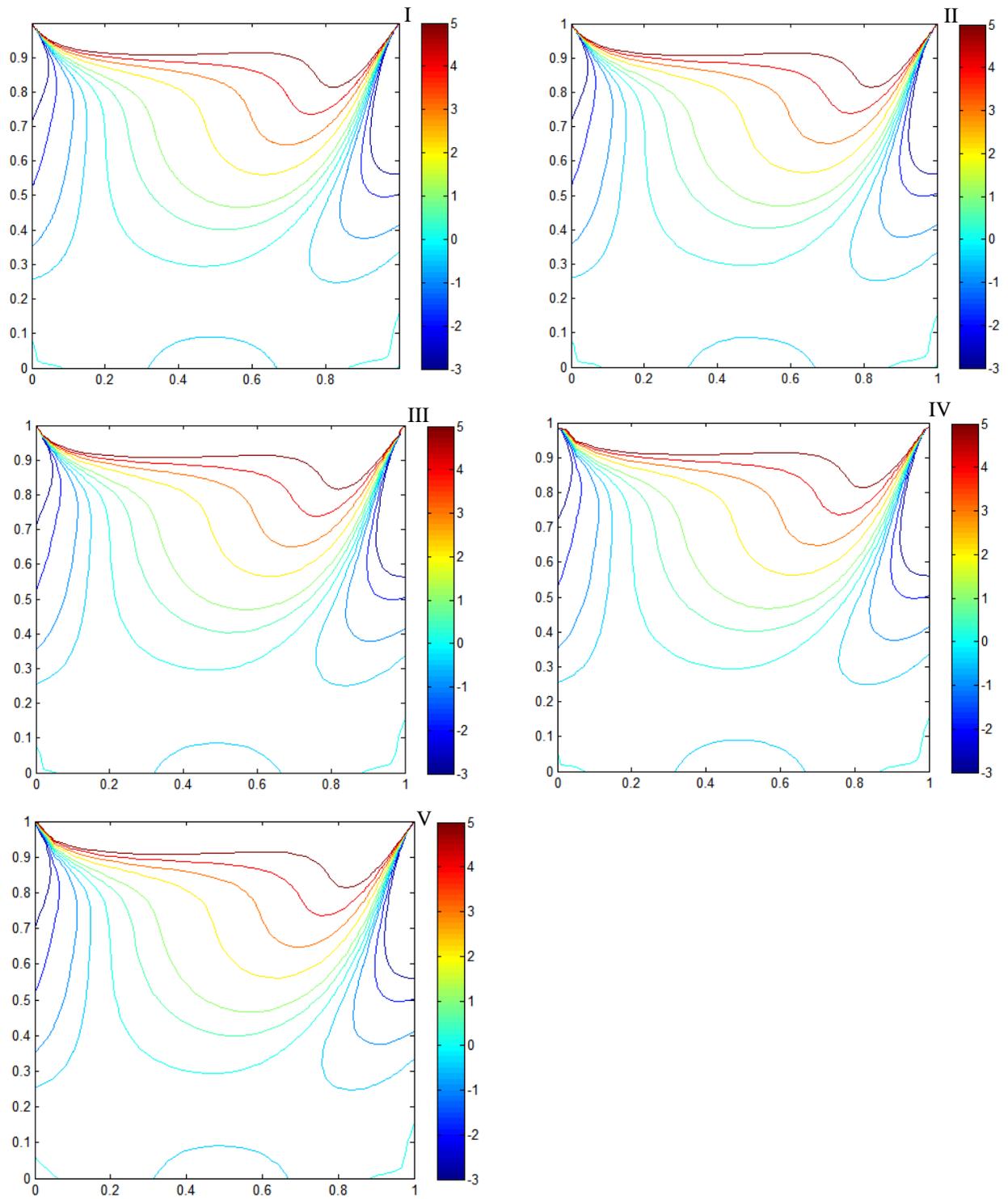

Fig. 14 Vorticity contour using different scheme shown in table 1

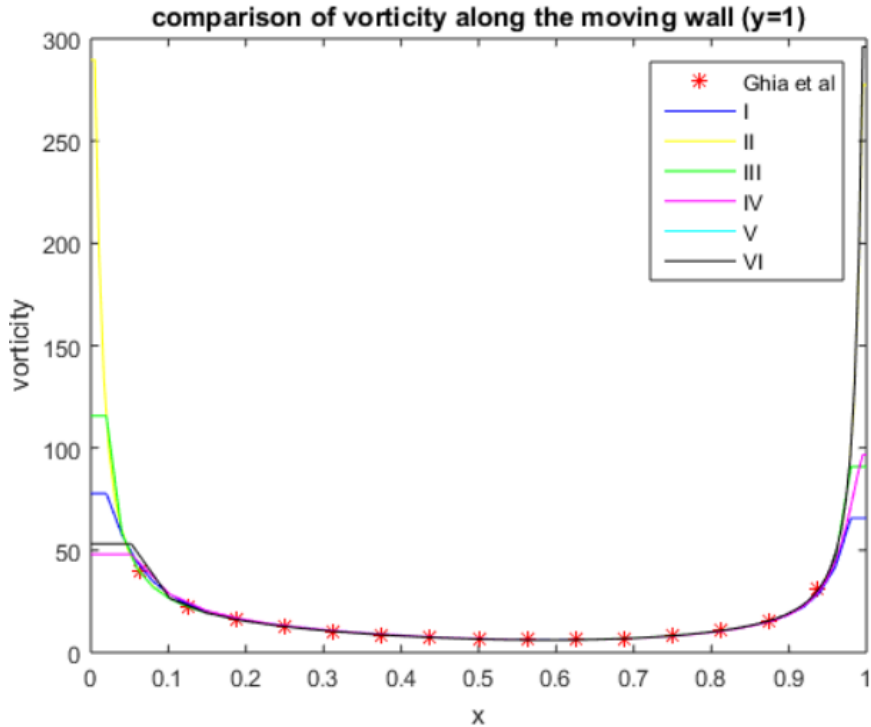

Fig. 15 Comparison of vorticity along the moving wall with Ghia's results

From figure 11, by comparing the results from 5 different schemes, one can easily conclude that the CFD method gives a good result on u-velocity in middle vertical line as grids are symmetric in x direction; similarly, from figure 12, we can also observe that the v-velocity in middle horizontal line matches well with the reference work as the grids are symmetric in y direction. That is, when the velocity in a certain direction is calculated, the numerical schemes using a symmetric grid in that given direction will perform better than those using non-symmetric grid. This is maybe because the irregular spacing ratio leads a worse accuracy in the CFD approximation using Tayler's expansion (the scheme with a uniform grid restores 2$^{nd}$ order accuracy in spatial while 1$^{st}$ order accuracy if non-uniform grids are adopted). And the error also results from the velocity approximation using stream function with an irregular spacing ratio. This can also explain why one will find that the vorticity and stream function own a better accuracy than the velocity if comparing figures 11 and 12 with figures 13 and 14.

From figure 13 and 14, one cannot see any significant difference of streamline and vorticity contour in the square area expect for the two bottom corners where the separation bubbles form. In general, the difference in the two corners mainly results from the density of elements. If more elements (or nodes) are adopted in that area, the numerical result will look better and smoother.

The same rule lies in figure 15 where the vorticity along the moving wall is compared using 5 different schemes with that from Ghia et al: the CFD scheme developed in this paper gives a good result in the middle area and the gap between these 5 schemes in the two ending domains also results from the grid density or how fine the mesh is. When a denser grid is adopted like scheme II in the left corner and scheme V in the right ending, there will gives a more accurate result. Otherwise, the numerical scheme will generate a relatively large error such as scheme IV in left and scheme I in right.

Overall, the numerical results obtained in this study show that the symmetric and fine mesh and grids lead to a better numerical solution, especially in the area where there exists a large gradient or divergence such as the four corners of cavity in this study.

## *4. Conclusion*

From the works list above, we can conclude that:

1. The FDM formulation developed in this paper with variable grid works well by comparing with Ghia's work;

2. The numerical scheme associated with a symmetric grid will generate a better solution in the whole domain than that with a non-symmetric grid.

3. The effect of the symmetry of grid on vorticity and stream line is not significant, and the vorticity and stream line mainly depends on the grid density especially in the corners: the denser grid gives a more accurate numerical result;

4. The symmetry and density of grids have a more significant influence on velocity than on vorticity and stream function because of double approximations on velocity using the variable grids.

5. A symmetric and fine mesh or grids should be used in the sensitive and interested area where there is a large gradient or divergence for a better accuracy while a relatively course and not totally symmetric mesh can be applied in the domain where you are less care about for a better efficiency.

# References

[1] Akshay Batra, Simulation of Lid Driven Cavity Problem using Incompressible Navier-Strokes Equation, MAE 561 CFD Final project, Arizona State University.

[2] Ghia, U., Ghia, K. N., & Shin, C. T. (1982). High-Re solutions for incompressible flow using the Navier-Stokes equations and a multigrid method. Journal of Computational Physics, 48(3), 387–411.

[3] Khorasanizade, S., & Sousa, J. M. M. (2014). A detailed study of lid-driven cavity flow at moderate Reynolds numbers using Incompressible SPH. International Journal for Numerical Methods in Fluids, 76(10), 653–668.

[4] Napolitano, M., & Pascazio, G. (1991). A numerical method for the vorticity-velocity Navier-Stokes equations in two and three dimensions. Computers and Fluids, 19(3–4), 489–495.

# Appendix

```matlab
clear all;
clc;

% This code is written by Zirui Mao to solve the 2D driven cavity problem
% using stream function - vorticity formulation as final project in EGFD 7051
@2016 Fall.

clear all
close all
M=51;
N=51;
x=zeros(M,N);
y=zeros(M,N);

% %%%%%%%%%%%%% Scheme VI %%%%%%%%%%%%%%%%%%%%
ratiox=1/1.2;  ]
ratioy=1/1.2;
deltax=(1-ratiox)/(1-ratiox^((M-1)));
deltay=(1-ratioy)/(1-ratioy^((N-1)));
for i=1:M
    x(i,:)=deltax*(1-ratiox^(i-1))/(1-ratiox);
end

for j=1:N
    y(:,j)=deltay*(1-ratioy^(j-1))/(1-ratioy);
end

%%%%%%%%%%%%% Scheme V %%%%%%%%%%%%%%%%%%%%
% ratiox=1/1.05;
% ratioy=1/1.05;
% deltax=(1-ratiox)/(1-ratiox^((M-1)));
% deltay=(1-ratioy)/(1-ratioy^((N-1)));
% for i=1:M
%     x(i,:)=deltax*(1-ratiox^(i-1))/(1-ratiox);
% end
%
% for j=1:N
%     y(:,j)=deltay*(1-ratioy^(j-1))/(1-ratioy);
% end

%%%%%%%%%%%%% Scheme IV %%%%%%%%%%%%%%%%%%%%
% ratiox=1/1.05;
% ratioy=1/1.1;
% deltax=(1-ratiox)/(1-ratiox^((M-1)));
% for i=1:M
%     x(i,:)=deltax*(1-ratiox^(i-1))/(1-ratiox);
% end
%
% for j=1:N
%     y(:,j)=(j-1)/(N-1);
% end
```

```matlab
% %%%%%%%%%%%%% Scheme III %%%%%%%%%%%%%%%%%%%%
% ratiox=1/1.1; % max_stretch^(1/((M-1)/2.0-1.0));
% ratioy=1/1.05; % max_stretch^(1/((N-1)/2.0-1.0));
% deltay=(1-ratioy)/(1-ratioy^((N-1)));
% for i=1:M
%         x(i,:)=(i-1)/(M-1);
% end
%
% for j=1:N
%     y(:,j)=deltay*(1-ratioy^(j-1))/(1-ratioy);
% end

%%%%%%%%%%%%% Scheme II %%%%%%%%%%%%%%%%%%%%
% ratiox=1.1;
% ratioy=1.1;
% deltax=(1-ratiox)/(1-ratiox^((M-1)/2.0))/2.0;
% deltay=(1-ratioy)/(1-ratioy^((N-1)/2.0))/2.0;
% for i=1:((M+1)/2.0)
%     x(i,:)=deltax*(1-ratiox^(i-1))/(1-ratiox);
% end
% for i=((M+1)/2.0+1):M
%     x(i,:)=deltax*(1-ratiox^((M-1)/2.0))/(1-ratiox)+deltax*ratiox^((M-1)/2.0-1.0)/...
%            (1-1/ratiox)*(1-(1/ratiox)^(i-(M+1)/2.0));
% end
%
% for j=1:((N+1)/2.0)
%     y(:,j)=deltay*(1-ratioy^(j-1))/(1-ratioy);
% end
% for j=((N+1)/2.0+1):N
%     y(:,j)=deltay*(1-ratioy^((N-1)/2.0))/(1-ratioy)+deltay*ratioy^((N-1)/2.0-1.0)/...
%            (1-1/ratioy)*(1-(1/ratioy)^(j-(N+1)/2.0));
% end

% %%%%%%%%%%scheme I%%%%%%%%%%%%%%%
% for i=1:M
%     for j=1:N
%         x(i,j)=(i-1)/(M-1);
%         y(i,j)=(j-1)/(N-1);
%     end
% end

% fid=fopen('Scheme I.txt','wt');% file path
% for i=1:1:M
%     for j=1:1:N
%             fprintf(fid,'%g\t%g\n',x(i,j),y(i,j));
%     end
% end
% fclose(fid);

U=1; % velocity in the top boundary [m/s]
dt=0.001; % time interval
writeInterval=50; %Timesteps
```

```matlab
dx=x(2:M,1)-x(1:M-1,1); % x interval vector
dy=y(1,2:N)-y(1,1:N-1); % y interval vector
nu=0.01; % 1/Re
vort=zeros(M,N);    % w
psi=zeros(M,N);     % psi
s=0;
t=0;
iter=0;
criterion_w=1.0e-15;
criterion_psi=1.0e-17;
Res_w=1.0;
Res_psi=1.0;

% initial condition settings
for i=1:M
    for j=1:N
        psi(i,j)=0;
        vort(i,j)=0;
        if j==M
            psi(i,j)=0;
            vort(i,j)=-2*U/dy(N-1);
        end
    end
end

while (Res_psi>criterion_psi && Res_w>criterion_w)
    vortold=vort;
    psiold=psi;
    iter=iter+1;
    t=t+dt;
    for i=1:M
        for j=1:N
            if j~=1 && j~=N && i~=1 && i~=M
                w_xx=2*(vort(i+1,j)*dx(i-1)+vort(i-1,j)*dx(i)-vort(i,j)*(dx(i)+dx(i-1)))/dx(i)/dx(i-1)/(dx(i)+dx(i-1));
                w_yy=2*(vort(i,j+1)*dy(j-1)+vort(i,j-1)*dy(j)-vort(i,j)*(dy(j)+dy(j-1)))/dy(j)/dy(j-1)/(dy(j)+dy(j-1));
                psi_x=(psi(i+1,j)-psi(i-1,j))/(dx(i)+dx(i-1));
                psi_y=(psi(i,j+1)-psi(i,j-1))/(dy(j)+dy(j-1));
                w_x=(vort(i+1,j)-vort(i-1,j))/(dx(i)+dx(i-1));
                w_y=(vort(i,j+1)-vort(i,j-1))/(dy(j)+dy(j-1));
                vort(i,j)=vort(i,j)+dt*(nu*(w_xx+w_yy)-psi_y*w_x+psi_x*w_y);
            elseif j==1 && i~=1 && i~=M
                vort(i,j)=2*(psi(i,j)-psi(i,j+1))/dy(1)^2; %Bottom
            elseif j==N && i~=1 && i~=M
                vort(i,j)=2*(psi(i,j)-psi(i,j-1))/dy(N-1)^2-2*U/dy(N-1); % Top
            elseif i==1 && j~=1 && j~=N
                vort(i,j)=2*(psi(i,j)-psi(i+1,j))/dx(1)^2; % Left
            elseif i==M && j~=1 && j~=N
                vort(i,j)=2*(psi(i,j)-psi(i-1,j))/dx(M-1)^2; % Right
            elseif (i==1 && j==1) %Set corners equal to neighbors or zero
                vort(i,j)=0;
            elseif (i==M && j==1)
                vort(i,j)=0;
            elseif (i==1 && j==N)
```

```matlab
                    vort(i,j)=(vort(i+1,j));
                elseif (i==M && j==N)
                    vort(i,j)=(vort(i-1,j));
                end
            end
        end

        for i=1:M
            for j=1:N
                if  j~=1 && j~=N && i~=1 && i~=M
                    Cx=2.0/dx(i)/dx(i-1)/(dx(i)+dx(i-1));
                    Cy=2.0/dy(j)/dy(j-1)/(dy(j)+dy(j-1));
                    psi(i,j)=(Cx*psi(i+1,j)*dx(i-1)+Cx*psi(i-1,j)*dx(i)+...
                        Cy*psi(i,j+1)*dy(j-1)+Cy*psi(i,j-1)*dy(j)+...
                        vort(i,j))/(Cx*(dx(i)+dx(i-1))+Cy*(dy(j)+dy(j-1)));
                else
                    psi(i,j)=0;
                end
            end
        end
        %calculate velocities
        for i=1:M
            for j=1:N
                if j~=1 && j~=N && i~=1 && i~=M
                    u(i,j)=(psi(i,j+1)-psi(i,j-1))/(dy(j)+dy(j-1));
                    v(i,j)=-(psi(i+1,j)-psi(i-1,j))/(dx(i)+dx(i-1));
                elseif j==N
                    u(i,j)=1;
                    v(i,j)=0;
                else
                    u(i,j)=0;
                    v(i,j)=0;
                end
            end
        end

        if iter/writeInterval == round(iter/writeInterval)
            s=s+1;
            times(s)=t;
            res_w(s)=max(max(abs(vort-vortold)));
            res_psi(s)=max(max(abs(psi-psiold)));
            Res_w=res_w(s);
            Res_psi=res_psi(s);
            fprintf('Calculating ... time = %f \n',t)

            set(0,'defaultfigurecolor','w')
            subplot(2,2,1)
            h=pcolor(x,y,u);
            set(h,'LineStyle','none')
            title('u');
            axis equal
            axis([0 1 0 1])
            xlabel({'x'},'FontSize',12)
            ylabel({'y'},'FontSize',12)
            colorbar;
```

```matlab
        colormap jet;
        
        subplot(2,2,2)
        h=pcolor(x,y,v);
        set(h,'LineStyle','none')
        title('v');
        axis equal
        axis([0 1 0 1])
        xlabel({'x'},'FontSize',12)
        ylabel({'y'},'FontSize',12)
        colorbar;
        colormap jet;
        
        subplot(2,2,3)
        h=pcolor(x',y',abs(vort'));
        set(h,'LineStyle','none')
        title('vorticity');
        axis equal
        axis([0 1 0 1])
        xlabel({'x'},'FontSize',12)
        ylabel({'y'},'FontSize',12)
        colorbar;
        colormap jet;
        
        subplot(2,2,4)
        h=pcolor(x',y',psi');
        set(h,'LineStyle','none')
        title('streamfunction');
        axis equal
        axis([0 1 0 1])
        xlabel({'x'},'FontSize',12)
        ylabel({'y'},'FontSize',12)
        colorbar;
        colormap jet;
        
        drawnow;
        
    end
    
end

figure;
set(0,'defaultfigurecolor','w')
plot(times,res_w,'r')
hold on
plot(times,res_psi,'b');
legend('residual vorticity','residual stream function');
xlabel('time')
ylabel('error')

% contour(x',y',psi',[10^-5 10^-6 10^-7 10^-8 -10^-10 -10^-7 -10^-5 -10^-4 -
0.0100 -0.03 -0.05 -0.07...
%           -0.09 -0.1 -0.11 -0.115 -0.1175 ]);
% colorbar
```

```matlab
%
% figure;
% contour(x',y',-vort',[-3.0 -2.0 -1.0 -0.5 0 0.5 1 2 3 4 5]);
% colorbar
%
%
% fid=fopen('velocity.txt','wt');% file path
% for i=1:1:M
%     for j=1:1:N
%             fprintf(fid,'%g\t%g\t%g\t%g\n',x(i,j),y(i,j),u(i,j),v(i,j));
%     end
% end
% fclose(fid);
%
% fidd=fopen('stream function and vorticity.txt','wt');% file path
% for i=1:1:M
%     for j=1:1:N
%
fprintf(fid,'%g\t%g\t%g\t%g\n',x(j,i),y(j,i),abs(vort(j,i)),psi(j,i));
%     end
% end
% fclose(fidd);
%
% fid1=fopen('vorticity in the top.txt','wt');% file path
% for i=1:1:M
%     fprintf(fid,'%g\t%g\n',x(i,N),abs(vort(i,N)));
% end
% fclose(fid1);
%
% fid2=fopen('velocity u along a vertical line passing through the center.txt','wt');% file path
% for j=1:1:N
%     fprintf(fid,'%g\t%g\n',y((M+1)/2.0,j),u((M+1)/2.0,j));
% end
% fclose(fid2);
%
% fid3=fopen('velocity v along a horizontal line passing through the center.txt','wt');% file path
% for i=1:1:M
%     fprintf(fid,'%g\t%g\n',x(i,(N+1)/2.0),v(i,(N+1)/2.0));
% end
% fclose(fid3);
```